\documentclass[prl,twocolumn,amsmath,amssymb,superscriptaddress,showpacs]{revtex4}
\usepackage{graphics}
\usepackage{epsfig}
\usepackage{bbm,times}

\newcommand{\tr}{{\rm Tr}}
\newcommand{\II}{\ensuremath{\mathbbm 1}}
\newcommand{\XX}{\mathcal X}
\newcommand{\YY}{\mathcal Y}
\newcommand{\ZZ}{\mathcal Z}
\newcommand{\NN}{\mathcal N}

\newcommand{\be}{\begin{equation}}
\newcommand{\ee}{\end{equation}}
\newcommand{\eea}{\end{eqnarray}}
\newcommand{\bea}{\begin{eqnarray}}

\newcommand{\mean}[1]{\ensuremath{\langle{#1}\rangle}}

\newcommand{\qed}{\ensuremath{\hfill \Box}}

\newcommand{\DD}{\ensuremath{\mathcal{D}}}
\newcommand{\BB}{\ensuremath{\mathcal{B}}}

\newcommand{\ketbra}[1]{\ensuremath{| #1 \rangle \langle #1 |}}

\newcommand{\ket}[1]{\ensuremath{|#1\rangle}}

\begin{document}
\title{Error-Tolerating Bell Inequalities via Graph States }
\date{\today}
\begin{abstract}
We investigate the Bell inequalities derived from the graph states
with violations detectable even with the presence of noises, which
generalizes the idea of error-correcting Bell inequalities [Phys.
Rev. Lett. {\bf 101}, 080501 (2008)]. Firstly we construct a family
of valid Bell inequalities tolerating arbitrary $t$-qubit errors
involving $3(t+1)$ qubits, e.g., 6 qubits suffice to tolerate single
qubit errors. Secondly we construct also a single-error-tolerating
Bell inequality with a violation that increases exponentially with
the number of qubits. Exhaustive computer search for optimal
error-tolerating Bell inequalities based on graph states on no more
than 10 qubits shows that our constructions are optimal for single-
and double-error tolerance.

\end{abstract}

\author{Qing Chen}

\affiliation{Hefei National Laboratory for Physical Sciences at
Microscale and Department of Modern Physics, University of Science
and Technology of China, Hefei 230026, P.R. China}

\affiliation{Centre for Quantum Technologies and Physics Department, National University of
Singapore, 2 Science Drive 3,  Singapore 117542}

\author{Sixia Yu}

\affiliation{Hefei National Laboratory for Physical Sciences at
Microscale and Department of Modern Physics, University of Science
and Technology of China, Hefei 230026, P.R. China}

\affiliation{Centre for Quantum Technologies and Physics Department, National University of
Singapore, 2 Science Drive 3,  Singapore 117542}

\author{C. H. Oh}

\affiliation{Centre for Quantum Technologies and Physics Department, National University of
Singapore, 2 Science Drive 3,  Singapore 117542}

\pacs { 03.67.Pp, 03.65.Ud, 03.67.Mn}

\maketitle

Quantum theory is inconsistent with local hidden variable (LHV)
theory, which is quantitatively characterized by the violations of
Bell inequalities \cite{bell64}. Experimentally the violations have
been confirmed only up to a certain extent \cite{loophole1,loophole2}. To close
the loopholes many efforts have been devoted to designing Bell
inequalities involving multi observers and multi measurement
settings in search for larger violations \cite{ghz,ghzbell,guhne,cabello}. Based
on the graph state \cite{graphs}, which is an essential resource in
the one-way computing \cite{1way}, the multi-observer Bell
inequalities are extensively studied \cite{guhne,cabello,divi,gbell1,gbell2}.

In general the Bell inequality is designed for some special
multipartite entangled quantum states, which may undergo some
inevitable errors. A powerful approach to fight the errors is to use
the quantum error-correcting codes. To protect our systems from
single qubit errors the simplest Bell inequality involves 10 qubits
by using the perfect 5-qubit code. In this case active decodings at
the detection steps are required. Recently an error-correcting Bell
inequality has been proposed based on some codewords of quantum
error-correction codes in which only passive detections are required
\cite{walker}. To tolerate  single qubit errors a minimum number of
11 qubits are involved in a valid Bell inequality.  Here we shall
employ the term {\em error-tolerating} instead of the original term
{\em error-correcting} because the violation can be detected without
involving any encoding-decoding procedures even when there are some
errors happened to the quantum state.

On the other hand the graph state turns out to be a systematic tool
for constructing good quantum codes, either additive or nonadditive,
binary or nonbinary \cite{grapherror,graphLC,yu1,yu2,cross,nonbin}.
It is therefore of much  interest to combine those two ideas: the
Bell inequalities from graph states and error tolerating to gain
some new insights. In this letter we firstly prove that every graph
state can be used to build an error-tolerating Bell operator, and
then by using some special graph states we build a valid
Bell inequality (with violations) on $3(t+1)$ qubits to tolerate up
to $t$-qubit errors. As a result, only 6 qubits are required instead
of the original 11 qubits \cite{walker} to tolerate single qubits
errors so that our error-tolerating Bell inequality can be possibly tested
under current experimental conditions \cite{exp,exp2}. Also we have
constructed a single-error tolerating Bell inequality with a
violation that increases exponentially with the number of the
qubits.

A graph $G=(V,E)$ is composed of a set $V$ of $n$ vertices and a set
of edges $E\subset V\times V$, i.e., two different vertices $a,b\in
V$ are connected iff $(a,b)\in E$. The neighborhood of a vertex $a$
is defined to be the set of all the vertices that are connected to
$a$, i.e., $N_a=\{b\in V|(a,b)\in E\}$. The graph state $\ket{G}$
corresponding to a graph $G$ on $n$ vertices is an $n$-qubit state
that is the unique joint +1 eigenstate of the following $n$
commuting observables
\begin{equation}\label{vs}
\mathcal G_a=\mathcal X_a\prod_{b\in N_a}\mathcal Z_b:=\mathcal
X_a\mathcal Z_{N_a}, \quad a\in V,
\end{equation}
which are referred to as {\it vertex-stabilizers}, i.e., $ \mathcal
G_a \ket{G}= \ket{G}, \mbox{ for } a=1,...,n$. Here $\mathcal
X_a,\mathcal Y_a, \mathcal Z_a$ are three Pauli operators acting on
the qubit $a$, and furthermore an operator subscripted by a subset stands for the
product of the same operator indexed by all the qubits in the
subset. For an arbitrary vertex subset $\omega\subseteq V$ the
observable $\mathcal G_\omega=\prod_{a\in \omega}\mathcal G_a$ also
stabilizes the graph state.

One distinct advantage of the graph state is that any Pauli operator
is equivalent to the product of a phase flip operator and a
stabilizer of the given graph state. In fact from Eq.~(\ref{vs}) it
follows that $ \ZZ_\delta\XX_\omega\propto\ZZ_{\delta\bigtriangleup
N_\omega}\mathcal G_\omega$ for arbitrary $\omega,\delta\subseteq
V$, where $N_\omega=\bigtriangleup_{v\in \omega}N_v$ is the
neighborhood of a subset $\omega$ with $C\bigtriangleup D:=C\cup
D-C\cap D$ being the \textit{symmetric difference} of any two sets
$C,D$. Thus all the nondegenerate Pauli operators acting on no more
than $t$ qubits
 will be equivalent to some phase flips
$\ZZ_C$ when acting on the graph state with $C\in \mathbb C_t$ where
\begin{equation}
\mathbb{C}_t=\left\{\delta\bigtriangleup
N_\omega\Big|\;|\omega\cup\delta|  \leq t\right\}
\end{equation}
is referred to as the $t$-coverable set.  In general we have
$\mathbb C_0=\{\emptyset\}$.  Based on the $t$-coverable set a
graphical approach has been developed to construct the quantum error
correction codes \cite{yu1,yu2,nonbin}.

\begin{figure}
\begin{center}\includegraphics[scale=1.0]{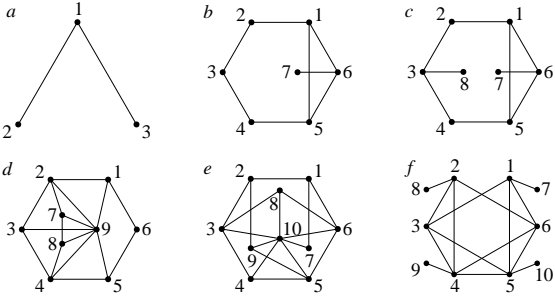}
\caption{ Some selected optimal graphs whose Bell operators
$\BB_t(G)$ have the largest violations: Figs.$1a, 1b \sim 1e$ for
$t=0$ and Fig.$1f$ for $t=1$.
 }
\end{center}
\end{figure}

For any given graph $G$ and the corresponding graph state
$|G\rangle$, since $\{\ZZ_C\mid C\in \mathbb C_t\}$ is exactly the
set of all the representative nondegenerate errors acting
nontrivially on no more than $t$ qubits, we introduce the
$t$-error-tolerating Bell operator in a similar manner as in Ref.
\cite{walker} \be \label{Berror} \BB_t(G)=\sum_{{C} \in
\mathbb{C}_t} \mathcal Z_C \ketbra{G} \mathcal Z_C^{\dagger}. \ee It
is obvious that $\BB_0(G)$ is exactly the Bell operator for the
3-setting Bell inequality constructed from the graph state
\cite{guhne}. For a simple example, we consider the 3-qubit GHZ
state corresponding to the star graph $\wedge$ as shown in Fig 1a.
We have $\mathbb C_t(\wedge)=2^V$ for $t\ge 1$ and as a result
 \bea
 8\BB_0(\wedge)=\II_1 \II_2 \II_3
+ \XX_1  \ZZ_2 \ZZ_3
+ \ZZ_1 \XX_2 \II_3
+ \ZZ_1 \II_2 \XX_3
\nonumber \\
+ \YY_1 \YY_2 \ZZ_3 + \YY_1 \ZZ_2 \YY_3 + \II_1 \XX_2 \XX_3 - \XX_1
\YY_2 \YY_3,
 \eea
while $8\BB_t(\wedge)=\II$ $(t=1,2,3)$. The following proposition
ensures that the expectation values of the Bell operator $\BB_t(G)$
under the corresponding graph state is error-tolerating.

{\it Proposition 1}: For a given graph $G$ on $n$ vertices with the
corresponding $n$-qubit graph state $\ket G$ and an arbitrary
trace-preserving completely positive map $\mathcal E_t$ described by
the Kraus operators $\{E_i\}_{i\in I}$ that are linear combinations
of Pauli operators nontrivially acting  on at most $t$ qubits,
 we have
 \be \label{Berror2n} \tr\Big(\BB_t(G)\mathcal E_t(\ketbra G)\Big)= 1.\ee

{\it Proof.} An arbitrary  Pauli operator acting nontrivially on no
more than $t$ qubits can be generally written as $\XX_{\omega}
\mathcal Z_{\delta}$ with $|\delta\cup\omega|\le t$ and, when acting
on a graph state $|G\rangle$, is proportional to a phase flip
$\mathcal Z_{C}$ up to a phase factor with $C=\delta \bigtriangleup
N_\omega\in \mathbb C_t(G)$. As a result we have expansion
 \be \label{Eifactor}
E_i=\sum_{C\in \mathbb{C}_t}\mathcal Z_{C}\sum_{\omega\in \Omega_C}
\lambda_{C, \omega}^i  \mathcal G_{\omega},
 \ee
where
$\Omega_C=\{\omega\subseteq V\mid \exists \delta\subseteq V,
|\delta\cup\omega|\le t\; s.t.\; \delta\bigtriangleup N_\omega=C\}$.
From the trace-preserving condition for $\mathcal E_t$, i.e., $
\sum_{i\in I} E_i^{\dagger} E_i=\II$, it follows that
 \be
\sum_{i\in I}\sum_{ C\in \mathbb C_t} \sum_{\omega,
\omega^\prime\in\Omega_C} (\lambda_{C, \omega^\prime}^{i })^*
\mathcal G_{\omega^\prime} \lambda_{C, \omega}^i \mathcal G_{\omega}
= \II
 \ee
which leads to
 \be \label{sumcoef} \sum_{i\in I}\sum_{
C\in\mathbb C_t}\left|\sum_{ \omega\in\Omega_C} \lambda_{C,
\omega}^{i }\right|^2= 1
 \ee
when averaged in the graph state $\ket
G$. As an immediate consequence
\begin{eqnarray}
\tr\Big(\BB_t(G)\mathcal E_t(\ketbra G)\Big)&=&\sum_{i\in I} \langle
G|E_i^\dagger\BB_t(G)E_i|G\rangle\cr &=& \sum_{i\in I,C\in \mathbb
C_t}\left|\langle G| \ZZ_C E_i|G\rangle\right|^2\cr
&=&\sum_{i\in I,C\in \mathbb C_t}\left|\sum_{ \omega\in\Omega_C}
\lambda_{C, \omega}^{i }\right|^2,
\end{eqnarray}
which yields the desired result.
 $\qed$

The above proposition shows that the expectation value of the Bell
operator $\BB_t(G)$ is the same when measured in the graph state
$|G\rangle$ no matter whether there are some errors acting on up to $t$ qubits or
not. That is why the Bell operator Eq.~(\ref{Berror}) is referred to as
error-tolerating. Next we shall investigate the maximal value of the
Bell operator $\BB_t(G)$ in the local hidden variable models in
order to have valid error-tolerating Bell inequalities.

For a given graph $G$ on $n$ vertices with the corresponding $n$-qubit
graph state $|G\rangle$, the Bell operator can be equivalently rewritten as
\be
\label{BGS}
\BB_t(G)=\frac1{2^n}\sum_{S\subseteq V}\sum_{C\in \mathbb
C_t}(-1)^{|C\cap S|}\mathcal G_S.
\ee
Now we have $n$ observers with each observer having 3
measurement settings corresponding to 3 Pauli operators
 $\mathcal X_a,\mathcal Y_a,$ and $\mathcal Z_a$ and they can
assume values $x_a, y_a, z_a =\pm 1$ independently.
By exhaustively calculating all possible realistic assignments,
we can determine the LHV bound
\be
\DD_t(G)=\max_{\rm LHV}\mean{\BB_t(G)}_c.
\ee
If for a given graph $G$ we have $\DD_t(G)<1$, then
 we have a valid Bell inequality, i.e., it can be violated. In this case the Bell inequality is
 error-tolerant because even there are up to $t$ arbitrary qubit errors,
if one measures the Bell operator $\BB_t(G)$ in the corresponding
graph state the violation can still be measured. The smaller the
$\DD_t(G)$ the larger the violation of the Bell inequality.

Since local clifford (LC) operations are special permutations
$\{\XX_i, \YY_i, \ZZ_i\} \rightarrow  \pm\{\XX_i, \YY_i, \ZZ_i\}$,
the
 LHV value $\DD_t(G)$ for the Bell operators defined on LC-equivalent graph states \cite{graphs,graphLC} should be the same.
For example, the complete graph and the star graph are
LC-equivalent, which correspond to the GHZ state, and therefore
their LHV bounds are equal. Starting from the GHZ state valid Bell
inequalities have been established when there is no error at all
\cite{ghzbell,guhne,walker}, i.e., for complete graph $K_n$ we have
$\DD_0(K_n)<1$ for $n>2$. However if some errors are permitted there
is no valid Bell inequality  that can be built from the  Bell
operator defined on $K_n$ since we have

{\it Proposition 2:} $\DD_t(K_n)=1$ for  $t\ge1$.

{\it Proof.}
Given a complete graph $K_n$, it is easy to obtain the $t$-coverable set for $t \geq 1$ as
\be
\mathbb C_t=\left\{ C \subseteq V \Big| |C| \leq t \text{  or  }  |C| \geq n-t \right\}.
\ee
It is clear that $C \in \mathbb C_t$ iff $V-C \in \mathbb C_t$ and because
$(-1)^{|C \cap S|}+ (-1)^{|(V-C) \cap S|}$ is zero when $|S|$ is odd we have in general
$\sum_{C\in \mathbb C_t} (-1)^{|C \cap S|}=0$ if $|S|$ is odd.
When $|S|$ is an even number we have
$\mathcal G_S=\YY_S$ for the complete graph (note that $\mathcal G_\emptyset=\YY_\emptyset=\II$ ).
For a given assignment of LHV values to $\YY_S$ let $A \subseteq V$ be the subset of qubits
on which $\YY_a$ is assigned to value -1 for all $a\in A$. Thus $\mathcal Y_S$ is assigned to the value $(-1)^{|S\cap A|}$ so
that the LHV value for the Bell operator reads
 \be
\langle {\BB_t(K_n)} \rangle =\frac1{2^n}\sum_{S\subseteq V}
\sum_{C\in \mathbb C_t}(-1)^{|S \cap (C \bigtriangleup A )|},
 \ee
which equals to $1$ if $A \in \mathbb C_t$ and $0$ otherwise.
 $\qed$

Instead of a single copy of the star graph we consider now two or
more copies of the star graph whose graph state is a direct product
of some GHZ states. Let  $\wedge^2=\wedge_1\oplus\wedge_2$ be the
graph on 6 vertices $V_1\cup V_2$ that is composed of two copies of
the star graph $\wedge$ on 3 vertices whose vertex sets are denoted
as $V_1$ and $V_2$. Its 1-coverable set is obviously $\{C|C\subseteq
V_1\mbox{ or } C\subseteq V_2\}$ from which the 1-error-tolerating
Bell operator can be calculated
 \be
\BB_1(\wedge^2)= \II \otimes
\BB_0(\wedge_2)+\BB_0(\wedge_1)\otimes(\II-\BB_0(\wedge_2)).
 \ee
It is easy to see that its LHV value $ \langle \BB_1(\wedge^2)
\rangle_c = b_2+ b_1 (1-b_2) $ with $b_i=\mean{\BB_0(\wedge_i)}_c\in
\{-1/4,1/4,3/4\}$ for $i=1,2$ reaches its maximum when
$b_1=b_2=3/4$. Therefore the LHV bound is $
\DD_1(\wedge^2)={15}/{16}<1$, which means that we have a valid Bell
inequality using 6 qubits instead of 11 qubits in \cite{walker} to
tolerate single-qubit errors.

As a direct generalization we consider the graph
$\wedge^m=\oplus_{i=1}^m\wedge_i$ on $3m$ vertices $\cup_{i=1}^m
V_i$ that composes of $m$ copies of the star graph $\wedge$ on 3
vertices whose vertices sets are $V_i$. The 1-coverable set for this
graph can be easily found to be $\{C|C\subseteq V_i \mbox{ for some
} i\}$ so that the 1-error-tolerating Bell operator can be
recursively expressed by
 \bea \BB_1(\wedge^{m+1})&=&\BB_1(\wedge^m)
\otimes \BB_0(\wedge_{m+1})\cr
&&+\BB_0(\wedge^m)\otimes(\II-\BB_0(\wedge_{m+1})).
 \eea
Because of the symmetry the recursive relationship above is the same
no matter which single copy is used for recurrence. In the following
we shall prove via induction
 \bea \label{D3m-1} &&\DD_1(\wedge^m)
=\left(1+\frac m3\right)\left(\frac34\right)^{m}.
 \eea
At first we notice the above LHV bound is attained when all the
variables are assigned to value +1, i.e., $b_i=\langle
\BB_0(\wedge_i) \rangle_c=3/4$ for all  $i\le m+1$, so that we have
only to prove that it is the upper bound of all LHV values. Suppose
that Eq.~(\ref{D3m-1}) holds true for $m$ copies. It is easy to check
the LHV value $\langle \BB_1(\wedge^{m+1}) \rangle_c$ when
$b_i=-1/4$ for all $i$ is smaller than the LHV value when $b_i=3/4$
for all $i$. Therefore, taking into account of the symmetry,  we can
suppose without lost of generality $b_{m+1}\ge0$ so that both
$b_{m+1}$ and $(1-b_{m+1})$ are nonnegative. By noticing
$\DD_0(\wedge^m)=(\frac34)^m\le \DD_1(\wedge^m)$ and $b_{m+1}\le
\frac 34$  we have
\begin{eqnarray}
\mean{B_1(\wedge^{m+1})}_c&\le&(1-b_{m+1})\DD_0(\wedge^m)+b_{m+1}\DD_1(\wedge^m)\cr& \le& \frac{3^{m}(m+4)}{4^{m+1}}.
\end{eqnarray}
Thus we have proved Eq.~(\ref{D3m-1}) for $m+1$. From this LHV bound
we see that the violation of the Bell inequalities increases
exponentially with the number of qubits.

On the same the graph $\wedge^m$ we consider the case $t=m-1$ in
which the $t$-coverable set can be easily found to be $\{C|C\cap
V_i=\emptyset\mbox{ for some } i\}$. The corresponding
$t$-error-tolerating Bell operator reads \be
\BB_{m-1}(\wedge^m)=\II-\bigotimes_{i=1}^m(\II-\BB_0(\wedge_i)), \ee
whose LHV value increases with the LHV values $\BB_0(\wedge_i)$ for
all $i=1,2,\ldots,m$. As a result we have the LHV bound \be
\DD_{m-1}(\wedge^m)=1-\frac1{4^m}. \ee The construction above can be
summarized as:

{\it Proposition 3:} There exists a valid $t$-error-tolerating Bell
inequality that involves only $3(t+1)$ qubits.

To build a valid Bell inequality, i.e., $\DD_t(G) < 1 $, it is
necessary that $\BB_t(G)$ not be the identity, i.e., the
$t$-coverable set $\mathbb C_t $ should not be the full set of all
the vertex subsets. This condition is necessary for the graph state
$\ket{G}$ being a base for some quantum error-correcting code that
correct up to $\lfloor \frac{t}{2}\rfloor$-qubit errors \cite{yu2}. However,
Proposition 3 shows that a valid error-tolerating Bell inequality is
not necessarily constructed from some $\lfloor
\frac{t}{2}\rfloor$-error-correcting codes. This is because by using
any copies of 3-qubit star graphs, only 1-error-correcting code can
be constructed.

For small $n$ it is possible to do a computer search on all the
$n$-qubit graph states for valid Bell inequalities. Before an
exhaustive computer search we notice that firstly for a given Bell
operator of form Eq.~(\ref{BGS}) we can restrict our attention to
those LHV models in which all $\ZZ$-measurements are assigned to
value +1. This is because \cite{guhne}  for a given vertex $v$, if we
revert the signs of the LHV values of all $\XX_a$ and $\YY_a$ in the
neighborhood of $v$, i.e., $a\in N_v$, and $\ZZ_v$ and $\YY_v$ then
LHV value of every term of the stabilizers $\mathcal G_s$ of the
graph state remains the same. And the Bell operator is a linear
combination of the stabilizers of the graph state.
Secondly the LHV bound for isomorphic graph and LC-equivalent graph is the same so that we
have to restrict to those non-isomorphic and non-LC-equivalent graphs.

\begin{table}
\caption{ The optimal values $\DD_t(n) $ for $3\le n\le 10$ and $0\le
t\le 2$ with corresponding graphs shown in Fig.1 (alphabetic labels)
and  explained in the text.}
\label{tab1}
\begin{tabular}{lcccccccc}
\hline\hline $t \setminus n$  & 3 & 4 & 5 & 6 & 7 & 8 & 9 & 10
\\
\hline 0& $3/4^a$ &3/4& 5/8&7/16&  $6/16^b$& $10/32^c$& $13/64^d$&$11/64^e$
\\
\hline 1& 1 &  1& 1 & $15/16^{\alpha}$ & $15/16^{\alpha}$ & $29/32^{\alpha}$  & $54/64^{\beta}$ & $48/64^f$
\\
\hline 2 &1  & 1& 1&1 &1 &1 & $63/64^{\beta}$ & $63/64^{\beta}$
\\
\hline\hline
\end{tabular}
\end{table}

We denote by $\DD_t(n)=\min_{\rm |V|=n} \DD_t(G)$ the minimal LHV
bound, i.e., largest violation, among all graphs on $n$ vertices and
its values for $n\le 10$ and $t\le 2$ are documented in Table I with
some of the optimal graphs shown in Fig.1. In the case of $t=0$ the
optimal values $\DD_0(n)$ for $n=3,4,5,6$ are attained by the ring
graphs as calculated in \cite{guhne} while our optimal values for
$n=7,8,9,10$, which are attained on the graphs shown in Fig.1b to
Fig.1e respectively,  improve the corresponding violations in
\cite{guhne}. Interestingly, in the case of $t>0$, many disconnected
graphs made up of complete graphs (or star graphs) can attain the
optimal bound. For examples, in Table I, those optimal values
labeled with $\alpha$ are attained by the graphs $K_3 \oplus
K_{n-3}$ $(n=6,7,8)$ and those values labeled with $\beta$ by graphs
$K_3 \oplus K_3 \oplus K_{n-6}$ $(n=9,10)$.
In the case of $t\ge 3$ we have always $\DD_t(n)=1$ for
$n\le 10$. It should be noted that the graphs attaining some of the
optimal values may not be unique and we have only listed one of the
optimal graphs.

If we define $\NN_t^{op} = \min \{n \big| \DD_t(n)<1 \}$ as the
smallest number of qubits that are involved in a valid
$t$-error-tolerating Bell inequality then the proposition 3
establishes an upper bound $\NN_t^{op}\le3(t+1)$. From Table I we
see that the upper bound is exact, i.e., the equality sign holds
true, in the case of $t=0,1,2$ and it is tempting to conjecture that
our upper bound is exact for any $t$.

In summary we have combined the idea of the Bell inequality via
graph states \cite{guhne} and the idea of the error-correcting Bell
inequalities \cite{walker} and gained some new sights. First of all a
$t$ error-tolerating Bell inequality is not necessarily built on
some $\lfloor\frac t2\rfloor$-error-correcting code and all the
graph states are possible candidates for the error-tolerating Bell
inequality. Secondly we have established the upper bound $3(t+1)$ of
the minimal number of qubits that are involved in  a valid
$t$-error-tolerating Bell inequality and this upper bound is exact
for $t\le 2$ as a result of an exhaustive computer search. It is
noteworthy that we have reduced the number of qubits from 11 to 6
that is involved in a 1-error-tolerating Bell inequality. Therefore
an experimental test is feasible \cite{exp,exp2}. Finally because the stabilizer
states are LC-equivalent to the graph states \cite{graphLC} our results hold in
fact for all the stabilizer states.

We acknowledge the financial support of NNSF of China (Grants  No.
10705025,  and No. 10675107) and the A*STAR Grant No.
R-144-000-189-305.

\end{document}